\documentclass[letter, twocolumn]{jpsj4}

\usepackage{graphicx}

\title{Superconductivity at 5.4 K in $\beta$-Bi$_2$Pd}

\author{Yoshinori Imai\thanks{E-mail address: imai@maeda1.c.u-tokyo.ac.jp}$^{1}$, Fuyuki Nabeshima$^{1}$, Taiki Yoshinaka$^{1}$, Kosuke Miyatani$^{1}$, Ryusuke Kondo$^{2}$, Seiki Komiya$^{3}$, Ichiro Tsukada$^{3}$, and Atsutaka Maeda$^{1}$
}

\def\runauthor{Y. Imai $et$ $al.$}

\inst{$^{1}$ Department of Basic Science, the University of Tokyo, 3-8-1 Komaba, Meguro-ku, Tokyo 153-8902, Japan\\
$^{2}$ Department of Physics, Okayama University, 3-1-1 Tsushimanaka, Kita-ku, Okayama, 700-8530 Japan\\
$^{3}$ Central Research Institute of Electric Power Industry, 2-6-1 Nagasaka, Yokosuka, Kanagawa 240-0196, Japan
}

\abst{We investigate bulk superconductivity in a high-quality single crystal of Bi$_2$Pd ($\beta$-Bi$_2$Pd; space group: $I4/mmm$) with a superconducting transition temperature of 5.4 K, by exploring its electrical resistivity, magnetic susceptibility, and specific heat.
The temperature dependence of the electrical resistivity shows convex-upward behavior at temperatures greater than 40--50 K, which can be explained using a parallel-resistor model. 
In addition, the temperature dependences of the upper critical magnetic field and the specific heat suggest that $\beta$-Bi$_2$Pd is a multiple-band/multiple-gap superconductor. }

\kword{superconductivity, multiple superconducting gaps, $\beta$-Bi$_2$Pd, Pd-Bi alloys, electrical resistivity, parallel-resistor model, upper critical field, specific heat}

\begin{document}
\maketitle


Studies of alloy superconductors (SCs) were of considerable interest in the 1950s and 1960s.  Matthias $et$ $al$. established the empirical law that the superconducting transition temperature, $T_\mathrm{c}$, depends on the number of valence electrons; this law is widely known as the Matthias rule\cite{RevModPhys.35.1}.
Among the Pd-Bi alloys, several superconducting materials, which were summarized in the review paper reported by Matthias $et$ $al$.\cite{RevModPhys.35.1}, have been identified: $\alpha$-BiPd (monoclinic structure, space group $P2_1$) with a $T_\mathrm{c}$ of 3.8 K; $\alpha$-Bi$_2$Pd (monoclinic structure, space group $C2/m$) with a $T_\mathrm{c}$ of 1.73 K; $\beta$-Bi$_2$Pd (tetragonal structure, space group $I4/mmm$) with a $T_\mathrm{c}$ of 4.25 K\cite{1957Zhu}; and $\gamma$-phase Pd$_{2.5}$Bi$_{1.5}$ (hexagonal structure, space group $P63/mmc$) with a $T_\mathrm{c}$ of 3.7--4 K\cite{1958Zhu}.
Among these alloys, the $\alpha$-BiPd phase has recently been investigated as a non-centrosymmetric SC\cite{PhysRevB.84.064518}.  The results of studies have shown that the anisotropy of $\alpha$-BiPd is not so large and that the overall effect of the no-inversion symmetry is of minor importance with respect to the bulk properties in $\alpha$-BiPd\cite{PhysRevB.84.064518}.  
However, no detailed reports concerning the physical properties of the other Pd-Bi superconducting phases, other than those that have detailed their $T_\mathrm{c}$ values and lattice parameters, have been published.

\begin{figure}[bt]
\begin{center}
\includegraphics[width=2cm]{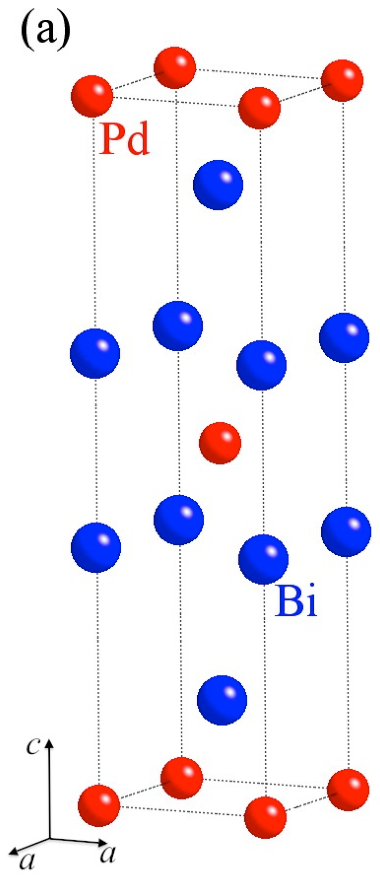}
\includegraphics[width=6cm]{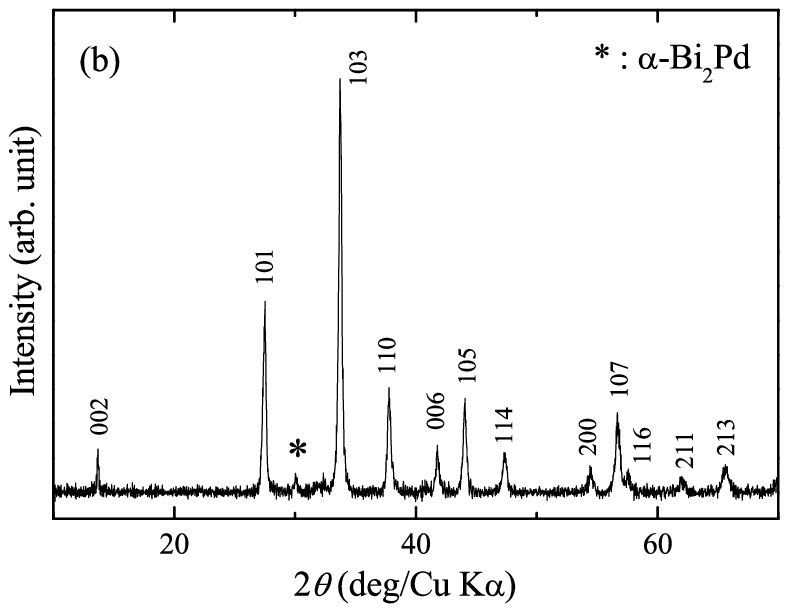}
\caption{
(color online) (a) Schematic crystal structure of $\beta$-Bi$_2$Pd.
(b) Powder X-ray diffraction pattern at room temperature using Cu K$\alpha$ radiation for $\beta$-Bi$_2$Pd single crystal.
  }
\label{fig:xrd}
\end{center}
\end{figure}%
%
%
In this letter, we focus on one of the Pd-Bi alloys, $\beta$-Bi$_2$Pd, the crystal structure of which is shown in Fig. \ref{fig:xrd}(a), and report the results of our investigations of a $\beta$-Bi$_2$Pd single crystal.
An early study\cite{1957Zhu} revealed that this compound showed superconductivity at temperatures less than 4.25 K.
However, we found that, by improving the crystal quality, the $T_\mathrm{c}$ of $\beta$-Bi$_2$Pd can reach 5.4 K.
In addition, the temperature dependences of the upper critical magnetic field and the specific heat suggest that $\beta$-Bi$_2$Pd is a multiple-band/multiple-gap SC. 
While multigap superconductivity, where the gaps on different parts of the Fermi surface become different magnitudes, was proposed theoretically\cite{PhysRevLett.3.552}, the first experimental observation of the possible existence of two distinct superconducting gaps was in the tunneling measurement of Nb-doped SrTiO$_3$\cite{PhysRevLett.45.1352}.
The existence of multiple supercoducting gaps leads to the anomalous temperature dependences of characteristics such as the specific heat, the upper critical magnetic field, and the penetration depth\cite{EPL.56.856, PhysRevLett.100.157001, RepProgPhys.71.116501, PhysRevB.66.180502, PhysRevB.78.024514, RevModPhys.83.1589, JPSJ.80.013704}.
After the discovery of the typical multigap SC, MgB$_2$\cite{Nat.410.63, EPL.56.856}, numerous studies on multigap superconductivity were carried out.
It is now well known that there are several multigap SCs such as NbSe$_2$\cite{PhysRevLett.90.117003}, Lu$_2$Fe$_3$Si$_5$\cite{PhysRevLett.100.157001, PhysRevB.78.024514}, and the iron-based SCs\cite{Kamihara08, RevModPhys.83.1589}.
One of the interesting aspects of multigap SCs is the variety of pairing mechanisms.
In iron-based SCs, the novel $s_\pm$-state, where a sign reversal of the gap function occurs between the hole and the electron pockets, has been proposed as a possible scenario\cite{PhysRevLett.101.057003, Kuroki08}. 
We demonstrate that $\beta$-Bi$_2$Pd is also a new candidate for a multigap SC, referring to the results of the specific heat and the upper critical magnetic field.

\begin{figure*}[hbt]
\begin{center}
\begin{minipage}{18cm}
\includegraphics[width=0.99\linewidth]{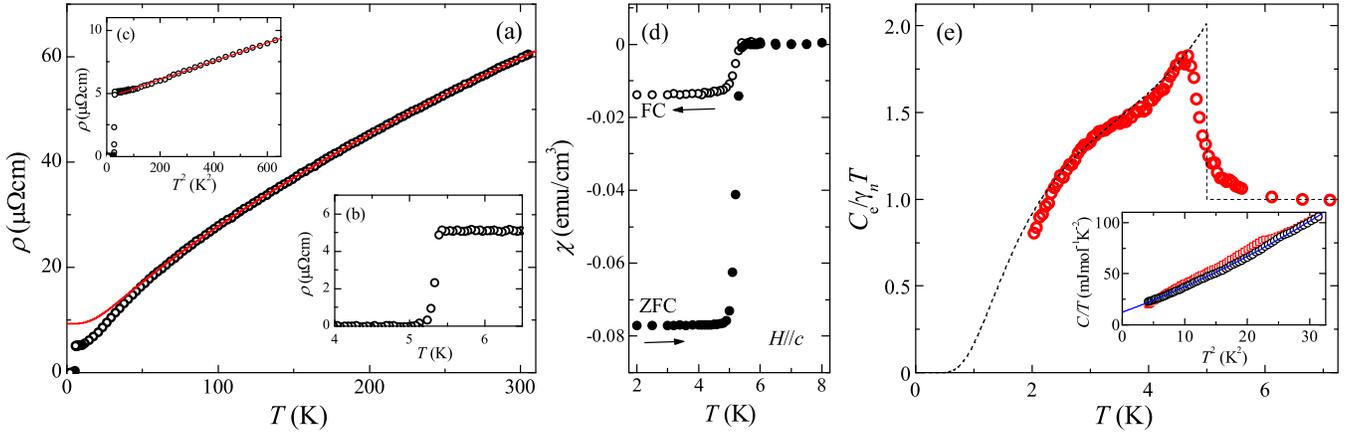}
\caption{
(color online) (a) Temperature dependence of the electrical resistivity ($\rho$) of a $\beta$-Bi$_2$Pd single crystal.
Inset (b) shows $\rho$ near $T_\mathrm{c}$; $\rho$ at  temperatures less than 25 K is plotted in inset (c) as a function of $T^2$.
(d) Temperature dependence of the magnetic susceptibility of a $\beta$-Bi$_2$Pd single crystal measured in a magnetic field of 2 Oe.  Closed and open circles represent the measurements in the zero-field cooling (ZFC) and field-cooling (FC) states, respectively.  
(e) Temperature dependence of normalized electronic specific heat in zero field.
The dashed curve is calculated using the two-band model ($2\Delta_1/k_\mathrm{B} T_\mathrm{c}=2.5$, $2\Delta_2/k_\mathrm{B} T_\mathrm{c}=6$, $\gamma_1/\gamma_n=0.90$), the details of which are described in the text.
The specific heat divided by temperature at $\mu_0 H= 0$ T (red squares) and 0.6 T (black circles) is plotted in the inset as a function of $T^2$.
The details of the blue solid curve are described in the text.
}
\label{fig:rhot}
\end{minipage}
\end{center}
\end{figure*}
%
%
Bi$_2$Pd single crystals were grown via a melt-growth method.   
The starting materials were Bi grains (5N) and a Pd wire (3N).  
These materials, in the prescribed molar ratio of Bi:Pd $=$ 2:1 (total: 2 g), were sealed in an evacuated quartz tube.
This quartz tube was heated at 900 $^\circ$C for 24 h, successively cooled to 600 $^\circ$C for 72 h, and then quenched in cold water.
All of the products were characterized by powder X-ray diffraction (XRD) using Cu K$\alpha$ radiation at room temperature.  Magnetic susceptibility measurements were performed using a superconducting quantum interference device (SQUID) magnetometer.  The electrical resistivity, $\rho$, was measured by the four-terminal method over the temperature range of 0.5 to 300 K under magnetic fields as strong as 3 T.
The specific heat was measured by the thermal-relaxation method at temperatures as low as 2 K on a commercial apparatus (Physical Property Measurement System, Quantum Design).

Figure \ref{fig:xrd}(b) shows the XRD pattern of a Bi$_2$Pd single crystal. 
Except for a few peaks that resulted from $\alpha$-Bi$_2$Pd, all of the peaks were indexed on the basis of a tetragonal lattice (no. 144, $I4/mmm$) with $a=3.37$ $\mathrm{\AA}$ and $c=12.96$ $\mathrm{\AA}$.  These lattice parameters are in good agreement with those reported previously\cite{1954Alek, 1957Zhu}.

The temperature dependence of $\rho$ for a $\beta$-Bi$_2$Pd single crystal is shown in Fig. \ref{fig:rhot}(a).
The large residual resistivity ratio [RRR, $\rho(T=300\ \mathrm{K})/\rho(T=6\ \mathrm{K})$] of 12 indicates the high quality of the crystal.
The $T_\mathrm{c}^\mathrm{onset}$, which is defined as the temperature at which $\rho$ begins to deviate from the normal-state behavior, and the $T_\mathrm{c}^\mathrm{zero}$, which is defined as the temperature at which $\rho$ becomes zero, were estimated to be 5.4 and 5.3 K, respectively, as shown in Fig. \ref{fig:rhot}(b).  These values are greater than the value of 4.25 K reported in previous papers\cite{1954Alek, 1957Zhu}.
The temperature dependence of the magnetic susceptibility in a magnetic field of 2 Oe is shown in Fig. \ref{fig:rhot}(d).
This result reveals that the diamagnetic transition of $\beta$-Bi$_2$Pd occurs at a temperature less than 5.4 K, which is in good agreement with the $\rho \left ( T \right )$ data.  
Here, it is interesting to note that the $T_\mathrm{c}$ of $\beta$-Bi$_2$Pd reported here is almost the same as that of Pd-intercalated Bi$_2$Te$_3$ with a very small superconducting volume fraction ($<1 \%$) in ref. 20, where the possibility that the topological insulator Bi$_2$Te$_3$ can be made into an SC by Pd intercalation between the Bi$_2$Te$_3$ layers is argued.

The temperature dependence of $\rho$ for $\beta$-Bi$_2$Pd exhibits the convex-upward characteristics at temperatures greater than 50 K; these characteristics are similar to those observed for A15 SCs\cite{PhysRev.136.A166, PhysRevB.13.5199, PhysRevLett.36.1084, RevModPhys.75.1085}.
Fisk and Webb have proposed that the resistivity of A15 compounds at high temperatures saturates at a value, $\rho_\mathrm{sat}$, that corresponds to the mean free path on the order of the interatomic spacing.\cite{PhysRevLett.36.1084}
Wiesmann $et$ $al.$\cite{PhysRevLett.38.782} developed the idea proposed by Fisk and Webb and found empirically that the $\rho$ of A15 compounds could be described using a parallel-resistor model:
\begin{equation}
\rho \left ( T \right ) = \left [ \frac{1}{\rho_\mathrm{sat}} + \frac{1}{\rho_\mathrm{ideal}\left ( T \right )} \right ]^{-1},
\label{para}
\end{equation}
where $\rho_\mathrm{sat}$ is the resistivity saturated at high temperature and is independent of $T$, and $\rho_\mathrm{ideal} \left ( T \right )$ is the ``ideal'' contribution according to Matthiessen's rule, $\rho_\mathrm{ideal} \left ( T \right ) = \rho_\mathrm{ideal, 0} + \rho_\mathrm{ideal, L} \left ( T \right )$.
Here, $\rho_\mathrm{ideal, 0}$ is the ideal temperature-independent residual resistivity caused by impurity scattering.
$\rho_\mathrm{ideal, L} \left ( T \right )$ is the temperature-dependent contribution caused by thermally excited phonons and can be expressed by the Bloch-Gr\"{u}neisen formula or by Wilson's theory:\cite{Ziman, WilsonTheory}\\
\begin{equation}
\rho_\mathrm{ideal, L} \left ( T \right )=C_1 \left ( \frac{T}{\theta_\mathrm{D}} \right )^r \int_{0}^{\frac{\theta_\mathrm{D}}{T}} \frac{x^r}{\left ( e^x-1 \right ) \left ( 1-e^{-x} \right )} dx,
\label{BlochGru}
\end{equation}
where $C_1$ is a numerical constant, $\theta_\mathrm{D}$ is the Debye temperature, and the values of the exponent $r$ are 3 and 5 for Wilson's theory and the Bloch-Gr\"{u}neisen formula, respectively.
The data for $\rho$ from 300 to 75 K were fitted to eq. (\ref{para}), and the fitted result is shown in Fig. \ref{fig:rhot}(a) as the solid curve.
For $\rho_\mathrm{ideal, L} \left ( T \right )$, we found that a better fit for $\rho \left ( T \right )$ in $\beta$-Bi$_2$Pd is given by Wilson's expression [specifically, $r=3$ in eq. (\ref{BlochGru})], which takes into account the interband electron-phonon Umklapp scattering between a low-mass s-band and a heavy-mass d-band\cite{WilsonTheory}.
The best-fitted result yields the values of 134 K for $\theta_\mathrm{D}$, 241 $\mu \Omega$cm for $\rho_\mathrm{sat}$, 9.63 $\mu \Omega$cm for $\rho_\mathrm{ideal, 0}$, and 63.3 $\mu \Omega \mathrm{K}^{-3}$ for $C_1$.
The value of $\theta_\mathrm{D}$ is very close to that obtained from the specific heat measurement, as will be discussed later. 
These results show that the parallel-resistor model explains the $\rho \left ( T \right )$ behavior of $\beta$-Bi$_2$Pd well at high temperatures.
In contrast, notable deviations between the experimental data and the parallel-resistor model are observed at low temperatures.  
In Fig. \ref{fig:rhot}(c), $\rho$ is plotted as a function of $T^2$ at low temperatures, which shows that the resistivity is proportional to $T^2$ at temperatures less than 25 K.
A similar crossover from the $T^2$ behavior to the saturated behavior upon heating has been observed in A15 compounds such as Nb$_3$Sn\cite{PhysRev.136.A166, PhysRevB.15.2624} and in $\beta$-pyrochlore oxides, $A$Os$_2$O$_6$ ($A=$K, Rb, Cs)\cite{JPSJ.81.011012}.
Some mechanisms of the $T^2$-dependence of $\rho \left ( T \right )$ have been proposed.\cite{PhysRev.135.A1333, PhysRevLett.56.647, Reiz1987, PhysRevB.56.10089, AdvPhys.33.257, PhysRevLett.99.187003}
However, the origin of the $T^2$-dependence of $\rho$ in $\beta$-Bi$_2$Pd cannot be specified solely from the results presented in this letter; further studies are needed.

Next, the specific heat divided by temperature, $C/T$, at $\mu_0 H = 0$ (red squares) and 0.6 T (black circles) is plotted in the inset of Fig. \ref{fig:rhot}(e) as a function of $T^2$.
$C/T$ at $\mu_0 H = 0.6$ T, where superconductivity is fully suppressed above 2 K, was fitted to the expression
\begin{equation}
C=\gamma_n T + \beta_n T^3 + \alpha_n T^5,
\label{sh}
\end{equation}
where $\gamma_n T$ is the electronic term, $C_\mathrm{e}$, and $\beta_n T^3 + \alpha_n T^5$ represents the phonon contribution.
From the fitting with eq. (\ref{sh}), which is shown in the inset of Fig. \ref{fig:rhot}(e) as the blue solid curve, we obtained the parameters $\gamma_n=12$ mJmol$^{-1}$K$^{-2}$, $\beta_n=2.3$ mJmol$^{-1}$K$^{-4}$, and $\alpha_n = 0.02$ mJmol$^{-1}$K$^{-6}$. 
The existence of the $T^5$ term in the normal-state specific heat suggests a complex phonon density of states.
From this value of $\beta_n$, $\theta_\mathrm{D}$ was estimated to be 136 K using the relation $\theta_\mathrm{D}=\left ( 12 \pi^4 N k_\mathrm{B}/5 \beta_n \right )^{1/3}$\cite{Ziman}, 
where $N$ is the number of atoms, and $k_\mathrm{B}$ is the Boltzmann constant.
This value of $\theta_\mathrm{D}$ is similar to that obtained from the analysis of the $\rho(T)$ data using eq. (\ref{para}), as previously mentioned.
The temperature dependence of normalized electronic specific heat at $\mu_0 H = 0$ T, which is estimated using the above parameters, is shown in Fig. \ref{fig:rhot}(e).
A clear jump appeared in $C_\mathrm{e}/\gamma T$ at a temperature of 5.0 K.  This value is slightly lower than the $T_\mathrm{c}$ estimated from the temperature dependences of $\rho$ and $\chi$.
The magnitude of the jump at $T=T_\mathrm{c}$, $\Delta C$, is 40 mJmol$^{-1}$K$^{-1}$, and the value of the normalized specific-heat jump, $\Delta C / \gamma_n T_\mathrm{c}$, is 0.82.
This value is smaller than that expected in the simple BCS weak-coupling limit, i.e., 1.43.
In addition, $C_\mathrm{e}$ of $\beta$-Bi$_2$Pd below $T_\mathrm{c}$ shows a peculiar temperature dependence.  That is, there is a plateau at approximately 3 K.
One might conclude that this plateau results from some impurity phases, for example, amorphous Bi or Bi-Pd alloys other than $\beta$-Bi$_2$Pd.  
However, there is no anomaly in $\chi (T)$ at $T \sim 3$ K.
Thus, it is unlikely that the origin of this plateau in the normalized electronic specific heat is an impurity phase.
These features, that is, the small jump at $T_\mathrm{c}$ and the plateau at approximately 3 K, in $C(T)$ of $\beta$-Bi$_2$Pd are familiar in the multigap SCs\cite{PhysC.355.179, EPL.56.856, PhysRevLett.100.157001}.
In the case of an SC with a single gap, the entropy, $S$, and $C$ are described as follows\cite{JLowTempPhys.12.387}:
\begin{equation}
\frac{S}{\gamma_n T_\mathrm{c}} = -\frac{6}{\pi^2k_\mathrm{B} T_\mathrm{c}} \int_{0}^{\infty}
 \left [  f \ln f + \left ( 1-f \right ) \ln \left ( 1-f \right ) \right ] d\epsilon,
\label{ent}
\end{equation}
\begin{equation}
\frac{C}{\gamma_n T_\mathrm{c}} = t\frac{d \left ( S/\gamma_n T_\mathrm{c} \right )}{dt},
\label{twogap}
\end{equation}
where $f=\left [ \exp \left ( E/k_\mathrm{B} T \right ) + 1 \right ]^{-1}$.
The energy of quasi particles is given by $E=\left [ \epsilon^2 + \Delta^2 \left ( t \right ) \right ]^{0.5}$, where $\epsilon$ is the energy of the normal electrons relative to the Fermi surface and $\Delta \left ( t \right ) = \Delta_0 \delta \left ( t \right )$ is the temperature dependence of the gap energy.
Here, $\delta \left ( t \right )$ is the normalized BCS gap at the reduced temperature, $t=T/T_\mathrm{c}$\cite{ZPhys.155.313}.
For the analysis of the data for $\beta$-Bi$_2$Pd, we use the two-band, two-gap model, where the total specific heat is considered as the sum of the contributions of each band calculated independently according to eq. (\ref{twogap}), as in the cases of MgB$_2$ and Lu$_2$Fe$_3$Si$_5$\cite{EPL.56.856, PhysRevLett.100.157001}.
Each band is characterized by the Sommerfeld coefficient, $\gamma_i$, with $\gamma_1 + \gamma_2 = \gamma_n$.
We calculate the specific heat by this two-gap model using three parameters of two gaps ($\Delta_1$, $\Delta_2$) and the relative weights ($\gamma_1/\gamma_n \equiv x$, $\gamma_2/\gamma_n \equiv 1-x$), and one of the calculated results is shown as the dashed curve in Fig. \ref{fig:rhot}(e).
The curve calculated using the two-gap model is in agreement with the experimental data, at least above 2 K, which suggests that $\beta$-Bi$_2$Pd is a multigap SC.
In this analysis, however, there is still some uncertainty and it is difficult to determine only one set of three parameters for lack of experimental data of $C$ at temperatures less than 2 K.
A more detailed analysis requires data at lower temperatures, and such measurements are currently in progress.

\begin{figure}[t]
\begin{center}
\includegraphics[width=7.5cm]{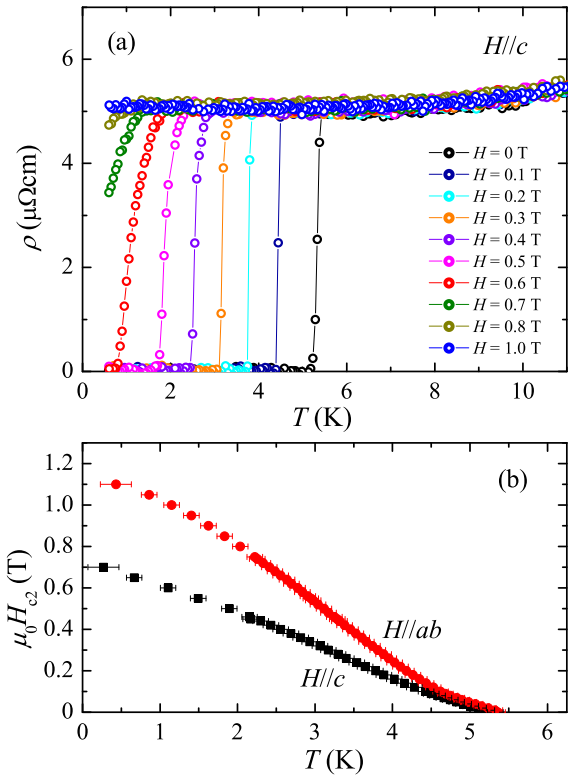}
\caption{
(color online) (a) Temperature dependence of the electrical resistivity of a $\beta$-Bi$_2$Pd single crystal in a magnetic field parallel to the $c$-axis.
The upper critical field, $\mu_0 H_\mathrm{c2}$, is plotted in (b) as a function of temperature.
Closed circles and squares represent the $\mu_0 H_\mathrm{c2}(T)$ data for magnetic fields parallel to the $ab$-plane and to the $c$-axis, respectively.
}
\label{fig:magrho}
\end{center}
\end{figure}
%
The effect of a magnetic field on $\rho$ is shown in Fig. \ref{fig:magrho}(a).
The $T_\mathrm{c}$ decreases almost linearly with increasing magnetic field.
The upper critical field, $\mu_0 H_\mathrm{c2}$, which is defined as the field in which $\rho$ becomes half the value of the normal-state resistance, is plotted in Fig. \ref{fig:magrho}(b) as a function of temperature.
The upper critical field extrapolated to $T=0$ K, namely, $\mu_0 H_\mathrm{c2}(0)$, is estimated to be $1.13\pm0.05$ T($H_\mathrm{c2}^{ab}(0)$) and $0.73\pm0.05$ T($H_\mathrm{c2}^{c}(0)$) for magnetic fields parallel and perpendicular to the $ab$-plane, respectively.
These results give Ginzburg-Landau coherence lengths of $\xi_{ab} (0) \sim 212\pm8$ $\mathrm{\AA}$ and $\xi_c (0) \sim 137\pm2$ $\mathrm{\AA}$, using $\mu_0 H_\mathrm{c2}^{ab}(0) = \Phi_0/2 \pi \xi_{ab}(0) \xi_c (0)$ and $H_\mathrm{c2}^{c}(0) = \Phi_0/2 \pi \xi_{ab} (0)^2$, where $\Phi_0 = 2 \pi \hbar / 2 e=2.07 \times 10^{-15}$ Tm$^2$ is the magnetic flux quantum.
The anisotropy parameter, $\Gamma$, which is defined as $\Gamma=H_\mathrm{c2}^{ab}(0)/H_\mathrm{c2}^{c}(0)$, is found to be 1.6.
It should be noted that the temperature dependence of $\mu_0 H_\mathrm{c2}$ in $\beta$-Bi$_2$Pd reveals a positive curvature close to $T_\mathrm{c}$, which becomes negative at temperatures less than approximately 3 K, as shown in Fig. \ref{fig:magrho}(b).
These temperature dependences have also appeared in other multigap SCs, such as MgB$_2$\cite{PhysRevB.66.180502, PhysicaC.456.160}, LaFeAs(O,F)\cite{Nat.453.903, RevModPhys.83.1589}, and SrPtAs\cite{JPSJ.80.055002}.
In addition, in some theoretical papers, this temperature dependence of $\mu_0 H_\mathrm{c2}$ has been explained on the basis of multiple superconducting gaps\cite{PhysRevB.67.184515, PhysRevB.69.054508, Kogan11}.
This result for $\mu_0 H_\mathrm{c2}$, together with the $C_\mathrm{e} \left ( T \right )$ data, suggests that $\beta$-Bi$_2$Pd is an SC with multiple superconducting gaps.  
Indeed, the presence of different Fermi surfaces has already been predicted by a band calculation\cite{JPhysCondMatt.4.2389}. 
Thus, our experimental findings suggest that the superconducting gaps open on different Fermi surfaces with different magnitudes.

In conclusion, we observed bulk superconductivity with a $T_\mathrm{c}$ of 5.4 K in $\beta$-Bi$_2$Pd by investigating the electrical resistivity, the magnetic susceptibility, and the specific heat.  The value of $T_\mathrm{c}$ reported in this letter is higher by approximately 1.2 K than those reported in previous papers and is the highest among the Pd-Bi alloy systems.
In addition, the temperature dependences of the upper critical field and the specific heat suggest that $\beta$-Bi$_2$Pd is a multigap superconductor.

\appendix
This work was performed using the facilities of the Cryogenic Research Center, the University of Tokyo, and was supported by a MEXT/JSPS Grant-in-Aid for Scientific Research (Grant Number 43244070).

\bibliographystyle{jpsj}

\providecommand{\noopsort}[1]{}\providecommand{\singleletter}[1]{#1}

\end{document}